\documentclass[3p]{elsarticle}

\usepackage{amsmath}
\usepackage{color}
\usepackage{eqnarray}
\usepackage{graphicx}
\usepackage{lineno,hyperref}
\usepackage{overpic}
\modulolinenumbers[5]

\journal{Journal of Fluids and Structures}

\biboptions{authoryear}

\newcommand\be{\begin{equation}}
\newcommand\ee{\end{equation}}

\newcommand{\tcr}[1]{\textcolor{black}{#1}} 

\def\Amean{{\left\langle A \right\rangle}}

\begin{document}

\begin{frontmatter}

\title{Stochastic modelling of a freely rotating disk facing a uniform flow} 

\author[eb]{Edouard Boujo\corref{cor1}} 
\address[eb]{LadHyX, UMR CNRS 7646, Ecole Polytechnique, 91128 Palaiseau, France}
\ead{edouard.boujo@ladhyx.polytechnique.fr}

\author[oc]{Olivier Cadot\corref{cor1}} 
\address[oc]{School of Engineering, The University of Liverpool, Liverpool L69 3BX, UK}
\ead{olivier.cadot@liverpool.ac.uk}

\cortext[cor1]{Corresponding author}

\begin{abstract}
The fluid-structure interaction between a thin circular disk and its turbulent wake is investigated experimentally and described with a low-order stochastic model.
The disk faces a uniform flow at Reynolds number $Re=$133 000 and can rotate around one of its diameters. It is equipped with instantaneous pressure measurements to give the aerodynamic loading; moments and centre of pressure of the front and base sides.
When the disk is fixed and aligned, 
symmetry-breaking vortex shedding is observed and all orientations are visited with equal probability, yielding axisymmetric long-term statistics.
Very-low-frequency antisymmetric modes as recently observed by \cite{Rigas_2014-JFM} are not observed unless the disk shape is modified into that of a bullet, showing the importance of the separation angle. It suggests that these modes are linearly stable for the disk while they are unstable for the bullet shape.
When the disk is fixed and inclined, the base center of pressure (CoP) shifts preferentially towards one side of the rotation axis. 
Finally, when the disk is free to rotate,
the front moment is proportional to the disk inclination (aerodynamic stiffness), while the base moment is well correlated with the disk inclination.
From these observations, a low-order model is derived that couples  
(i)~the CoP, forced stochastically by the turbulent flow, and
(ii)~the disk inclination, as a linear harmonic oscillator forced by the base moment.
The model captures well the main statistics and dynamic features observed experimentally. One difference lies in the distribution of the disk inclination (which features exponential tails in the measurements but is Gaussian in the model), suggesting a possible refinement with non-linearities or time delays.
The model is expected to capture coupled dynamics in other systems involving turbulent wakes behind freely rotating bluff bodies.
\end{abstract}


\end{frontmatter}


\section{Introduction}
Turbulent dynamics of massively separated wakes produced by simple axisymmetric bluff bodies such as a sphere or a disk past a uniform flow have always been a central problem in fundamental fluid mechanics. The flow complexity in space associated with unpredictable temporal fluctuations make both the experimental and numerical approaches difficult. The first case is limited by the available measurements and the second suffers from too small duration of computation. Early studies provided extensive characterizations of periodic or quasi-periodic phenomena in the wake \citep{Achenbach-1974,Taneda-1978,fuchs_mercker_michel_1979,Sakamoto-1990,Berger-1990,Yun-2006}, but for instance all attempts to describe the topology of vortex shedding related to the antisymmetric mode never met any clear consensus.

Recently, \cite{Rigas_2015-JFM} proposed a stochastic wake model based on observations obtained from 
\tcr{pressure distribution measurements on the base of an axisymmetric body}
and velocity measurements in a cross flow plane \citep{Rigas_2014-JFM, Grandemange_2014-ExiF,Gentile-2016}. When adapted to the centre of pressure dynamics expressed in radial and phase angle coordinates, the model is able to restore satisfactorily the diffusive behaviour of the phase angle as well as the permanent wake asymmetry reminiscent of the first steady bifurcation in the laminar regime \tcr{\citep{Fabre_2008,Meliga_2009}}. Body misalignment, studied by \cite{Gentile-2017}, confines the centroid of the back-flow region of the wake around an average off-centre position, equivalently to a constraint of the phase angle exploration of the centre of pressure. \cite{Cadot_2016_jfmrapids} investigated the interaction of a rectangular flat plate, free to rotate about the minor axis, with the wake dynamics by means of flow visualization techniques in water. The fluid structure coupling revealed angle time series with long-time stochastic dynamics with no periodicity for plates aspect ratio of order 1.

The objective of the paper is to model such fluid-structure interaction produced by three-dimensional massive flow separation on the basis of the stochastic model of \cite{Rigas_2015-JFM}. To conserve the axisymmetric wake framework, we will consider a disk facing an air flow. The disk is free to rotate about an axis passing through its diameter. Quantitative information will be experimentally provided by the measurements of the instantaneous pressure distribution at the disk surface.

The paper is organized as follows.
Section~\ref{sec:expe} describes the experimental set-up.
Section~\ref{sec:fixedmodel} presents measurements for the fixed disk (aligned or inclined), and a stochastic model for the dynamics of the  center of 
\tcr{pressure on the base side}, including the effect of the disk inclination.
Section~\ref{sec:FSImodel} presents measurements for the freely rotating disk, and the stochastic model is extended to describe the dynamics of the disk inclination forced by the base moment.
Simulation results of the model are commented in
Sec.~\ref{sec:results}.
Conclusions are drawn in Sec.~\ref{sec:conclusion}.

\section{Experimental set-up}
\label{sec:expe}

\begin{figure}[h]
\setlength{\unitlength}{1cm}
\begin{center}
\includegraphics[width=15cm]{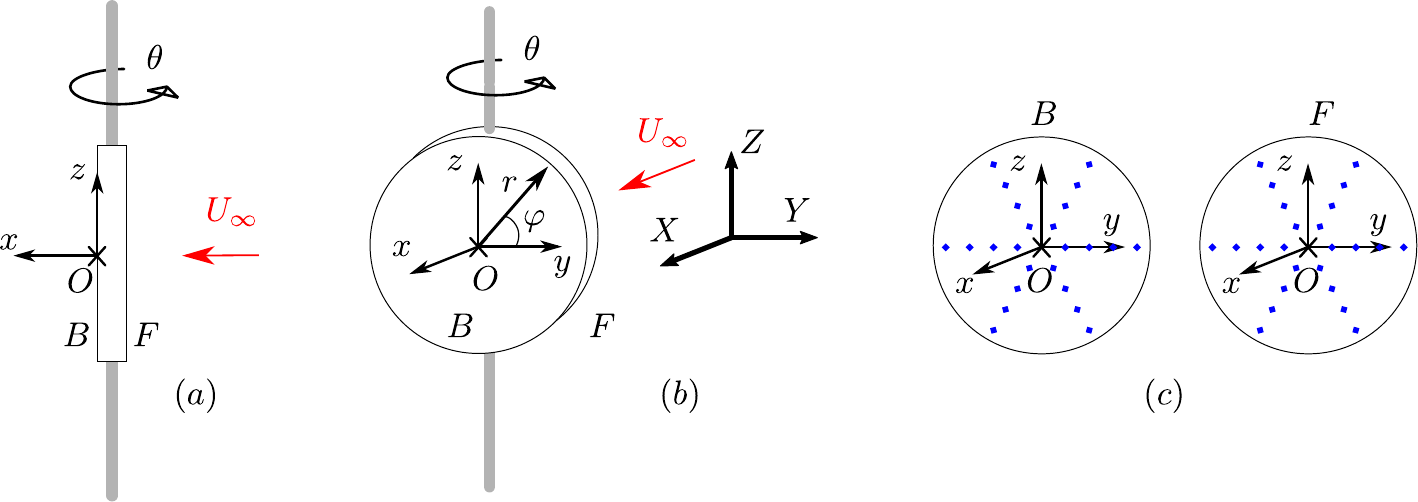}
\end{center}
\vspace{-0.5cm}
\caption{Side ($a$) and rear ($b$) views of the experimental geometry. Letters $B$ and $F$ respectively stands for the base and the front side. 
The frame ($Oxyz$) is associated with the disk and in the laboratory frame ($XYZ$), $X$ corresponds to the main flow direction $U_\infty$. 
Blue symbols in ($c$) represent the locations of the pressure measurements, 24 taps on each disk side. Note that in ($c$) the front side is seen from behind.}
\label{fig:experiment}
\end{figure}

\tcr{A disk of diameter $D=20$~cm and thickness $e=2.5$~cm is equipped with a tube of diameter $1$~cm passing through the disk centre as depicted in Fig.~\ref{fig:experiment}. The disk is hollow and built from $3$~mm Plexiglas sheets to allow wall mounted pressure tapping. The tube is mounted on two ball bearings at the floor and the roof of the wind tunnel to ensure the free rotation of the system.} The disk is placed at mid-height of the test section, which  is $60$~cm high and $120$~cm wide. A laser displacement sensor from Keyence is used to measure the deflection angle $\theta$. Both sides of the disk are equipped with 24 pressure measurements displayed in Fig.~\ref{fig:experiment}($c$). They are $0.8$~mm holes in diameter and connected via vinyl tubes to a differential pressure scanner, a model ZOC~33 from Scanivalve. The vinyl tubes are passed through the axis of rotation and the pressure scanner is placed outside the wind tunnel. 
The $2.5$~m length of each vinyl tube introduces a low-pass filtering with a cut-off frequency estimated at $f_C\simeq 15$~Hz from spectra analysis (i.e. $\frac{f_C D}{U_\infty}=0.3$ in non-dimensional units). The spatial pressure distribution is obtained from $6$ rows (at azimuthal angles $\varphi= n \frac{\pi}{3}$) of 4 holes (with radial positions of $r=3,5,7,9$~cm). 
Two channels of the pressure scanner also measure the static pressure $p_\infty$ and dynamic pressure $q$ taken upstream with a Pitot tube.
The discrete radial pressure distribution $p(r,\varphi)$ is interpolated at $r=0,1,...,10$~cm using a cubic spline method. A $3/8$ Simpson integration rule is used for the computation of the aerodynamic moments about the axis of rotation $(Oz)$,  normal forces and the centre of pressure (CoP) produced by each pressure distribution of the front and base side. 
\tcr{We refer to pressure on the front and base sides as ``front pressure'' and ``base pressure''.}
In the frame ($Oxyz$) associated with the disk such that ($Ox$) is the outward normal of the base as depicted in Fig.~\ref{fig:experiment}($b$), 
the expressions of the moments and forces are respectively:
\begin{eqnarray}
M^i=s_i\int\!\!\!\int_i y \, (p-p_\infty)r \,\mathrm{d}r \mathrm{d}\varphi,
\quad F_N^i=-s_i\int\!\!\!\int_i (p-p_\infty)r \,\mathrm{d}r \mathrm{d}\varphi;
\label{eq:def_moment_force_cop}
\end{eqnarray}
where 
$i=B$ (base) or $i=F$ (front), and
$s_B=+1$ and $s_F=-1$ due to the opposite side orientations. 
Force coefficients $C_N^i$ and moment coefficients $C_M^i$ are defined in the usual way:
\begin{eqnarray}
C_N^i=\frac{F_N^i}{qS}, \quad
C_M^i=\frac{M^i}{qSD},
\end{eqnarray}
where $S=\pi D^2/4$ is the disk surface and $q=\frac{1}{2}\rho_f U_\infty^2$.

The working fluid is air and the incoming flow speed is set to $U_\infty=10$~m.s$^{-1}$. The flow Reynolds number based on the disk diameter is then $Re=\frac{U_\infty D }{\nu} \simeq 1.33 \times 10^5$. 
The pressure signals (i.e. the aerodynamics loading) and the disk angle $\theta$ are recorded at the sampling frequency of $500$~Hz for 40 minutes to ensure appropriate convergence of the statistics.

\section{Flow model for the fixed disk}
\label{sec:fixedmodel}

We first consider the disk at fixed angle positions to investigate the wake dynamics without fluid-structure interaction. 
The pressure distribution measurements on both sides of the disk will be used to produce an adequate model for the flow.

\subsection{Base loading}
\label{sec:base}

Azimuthal modes on the base of the disk are extracted using the same methodology as in \cite{Rigas_2014-JFM}. 
First, each crown of pressure measurements at instant time $t$ is decomposed in Fourier components:
\begin{eqnarray}
\tilde p_m (r,t)=\frac{1}{2\pi} \int_0^{2\pi} p(r,\varphi,t) e^{-im\varphi} \,\mathrm{d}\varphi.
\end{eqnarray}
Coherent structures are identified by calculating the azimuthal spectral energy density $\Phi_m(St)$ that is distributed over Strouhal number $St=f D/U_\infty$ (non-dimensional frequency) for each crown:
\begin{eqnarray}
 \Phi_m (r,St) = \left| \frac{1}{2\pi}\int_0^{\infty}\tilde p_m(r,t^*)e^{-2\pi i St~t^*} \,\mathrm{d}t^* \right|^2,
\end{eqnarray}
and averaged over the radius such that 
\begin{eqnarray}
|\tilde p_m |^2 = \int_0^\infty\frac{8}{D^2}\int_0^{D/2}\Phi_m(r, St)  r \,\mathrm{d}r \mathrm{d}St 
= \int_0^\infty \bar\Phi_m(St) \,\mathrm{d}St.
\end{eqnarray}

\begin{figure}
\setlength{\unitlength}{1cm}
\begin{center}
\includegraphics[height=8cm]{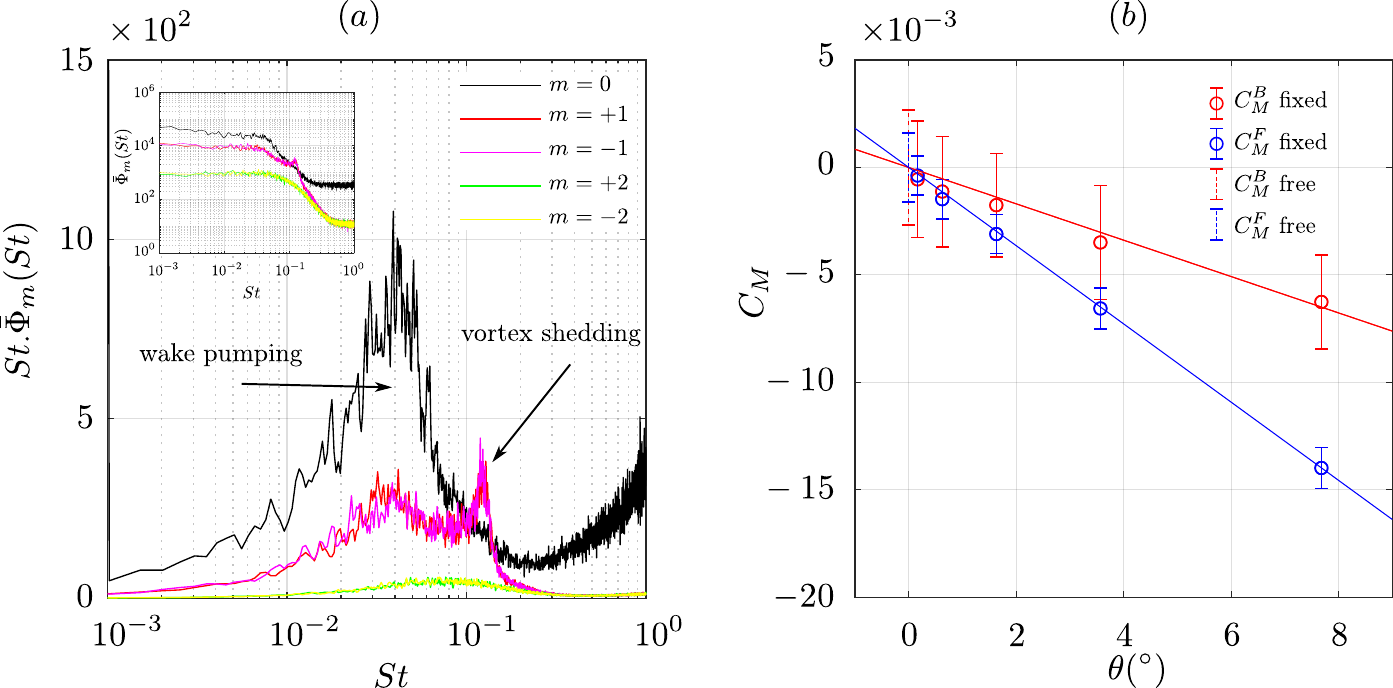}
\end{center}
\vspace{-0.5cm}
\caption{Premultiplied power spectra (inset: power spectra) of the azimuthal modes on the base of the fixed disk at $\theta = 0^\circ$. Mean moment coefficient ($b$) measured with the fixed disk (symbols and solid lines) and the free disk (dashed lines) produced by the base and front side of the disk. Each vertical line has the standard deviation in length.}
\label{fig:pressure-spectra}
\end{figure}

The averaged spectral density $\bar\Phi_m(St)$ (computed over 60 windows of $20 000$ points at $500~$Hz, thus giving a frequency resolution of $\mathrm{d}f=0.025~$Hz or $\mathrm{d} St=5\times 10^{-4}$) is shown for the disk in Fig.~\ref{fig:pressure-spectra}$(a)$. 
The premultiplied spectrum is of practical interest because the energy can be estimated quickly (linearly) with the area under the curve.
The base pressure dynamics is dominated by the antisymmetric periodic modes $m=\pm1$ (vortex shedding) around the Strouhal number of 0.125. At lower frequencies, the axisymmetric mode $m=0$ is definitely dominating the dynamics. 
The high-frequency plateau of the mode $m=0$ above $St=0.2$ in Fig.~\ref{fig:pressure-spectra}($a$) is likely to be due to the wind tunnel acoustics noise level, \tcr{because it corresponds to in-phase fluctuations of the 48 pressure taps.} Below this frequency, the energy corresponds to the breathing or pumping dynamics of the recirculating bubble with a maximum of energy around $St=0.04-0.06$ as shown by the premultiplied spectra. 
In contrast to the results about axisymmetric wakes of bullet shapes from \cite{Grandemange_2014-ExiF,Rigas_2014-JFM,Rigas_2015-JFM,Rigas_2016,Gentile-2016,Gentile-2017} there is no very-low-frequency antisymmetric $m=\pm 1$  mode for $St<0.01$. To double-check this important and surprising result, 
we added a hemispheric shape on the forebody side to reproduce separation angles similar to those of bullet shapes. 
The very-low-frequency antisymmetric azimuthal mode are retrieved, similarly to the primary work of \cite{Rigas_2014-JFM}.
\tcr{The presence or absence of these modes in the base pressure distribution is of fundamental importance and further investigations, beyond the scope for the present study, are required to understand the cause.}

\subsection{Front loading}
The flow facing the disk is much less unsteady than that of the base as can be seen in Fig.~\ref{fig:pressure-spectra}$(b)$ by comparing both standard deviations of the moment coefficient $C_{M}^i$ ($i=B,F$) computed on each side and displayed as  error bars. 
To define the full alignment of the disk with the flow $(\theta=0)$, the free disk mean angle and the corresponding mean moment coefficients ($C_{M}^B=-8\times 10 ^{-5}$ and $C_{M}^F=-1.45\times 10 ^{-3}$) are taken as zeros. 
The discrepancy to expected null quantities is produced by the mean flow divergence, the accumulation of geometrical imperfections and the uncertainties in the pressure measurements. 
Figure~\ref{fig:pressure-spectra}$(b)$ shows that the flow produces a mean restoring moment linear in the angle deflection defining the aerodynamic stiffnesses. 
The front side loading of the fixed disk is thus mainly due to the restoring moment:
\begin{eqnarray}
C_{M}^F=-\frac{k}{qSD}\theta=-1.82\times10^{-3}(^{\circ^{-1}})\theta
=-0.104\,\theta.
\label{eq:hydro_stiff}
\end{eqnarray}
In the following, the effect of the front loading on fluid-structure interaction will be modelled via this aerodynamic stiffness.

\subsection{Model of the wake deflection}

\begin{figure}
\setlength{\unitlength}{1cm}
\begin{center}
   \begin{overpic}[height=7.5cm, trim=15mm 92mm 30mm 70mm, clip=true]{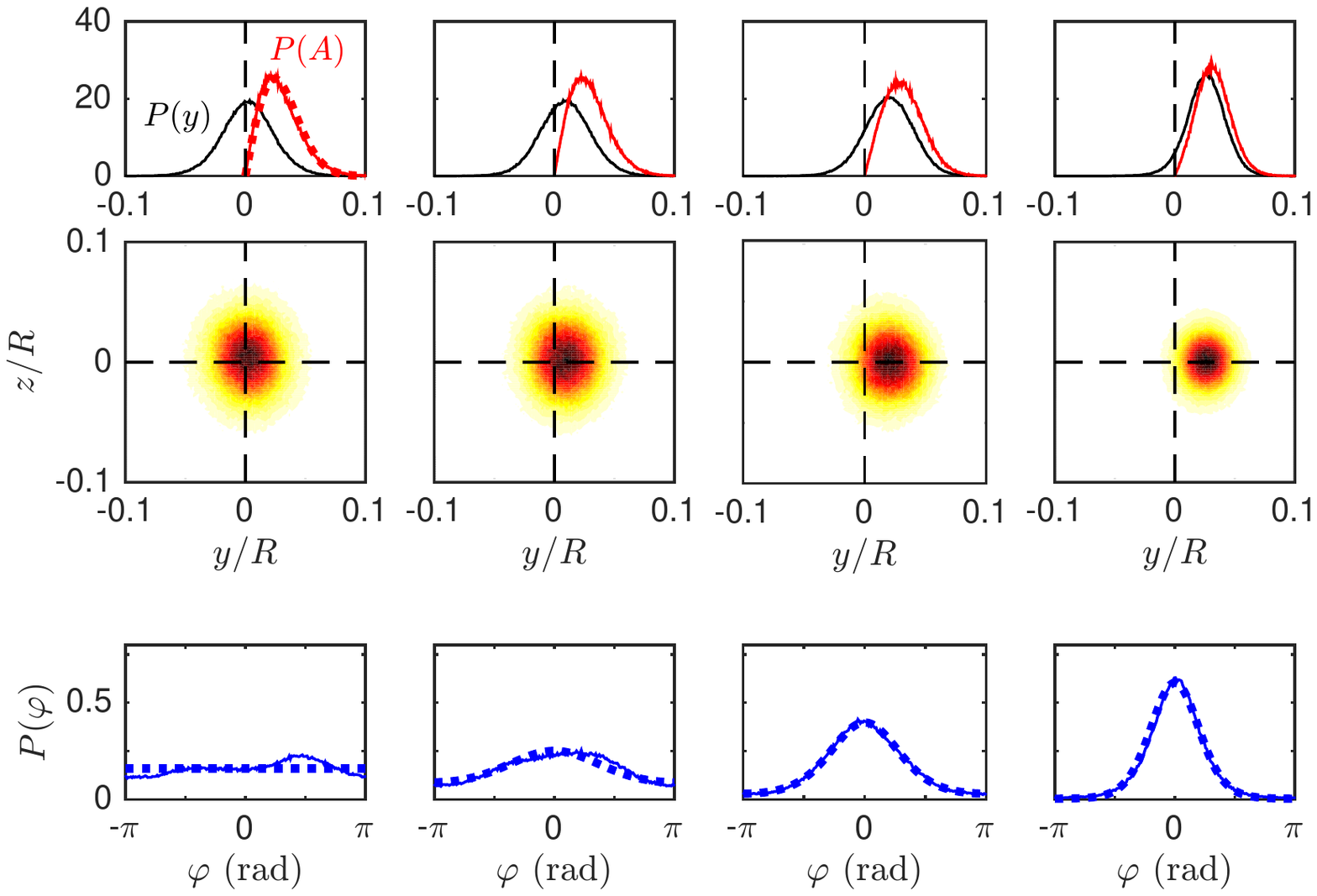}
      \put(15.5,69){$\theta=0^o$}
      \put(38.5,69){$\theta=2^o$}
      \put(61.5,69){$\theta=5^o$}
      \put(85.1,69){$\theta=10^o$}
      \put(-5,64){$(a)$}
      \put(-5,47){$(b)$}
      \put(-5,17){$(c)$}
   \end{overpic}
\end{center}
\vspace{-0.5cm}
\caption{
Distribution of the base center of pressure for increasing fixed disk inclination.
$(a)$~PDF of the CoP $y$ location and CoP radius $A$.
$(b)$~Joint PDF of the CoP $y$ and $z$ locations.
$(c)$~PDF of the CoP orientation $\varphi$.
In $(a)$ and $(c)$: (---) measurements and  $(\cdots)$  
stochastic model~(\ref{eq:A_Phi_th}).
}
\label{fig:PDF_2D_CoP}
\end{figure}

We come back to the base side and take a closer look at the wake's stochastic dynamics.  
To this aim, we use the base CoP location, normalized by the disk radius $R=D/2$:
\be 
\mathbf{CoP}^B = \frac{1}{R}
\frac{\int\!\!\!\int_i 
\mathbf{r} \, (p-p_\infty)r 
\,\mathrm{d}r \mathrm{d}\varphi}
{\int\!\!\!\int_i~ (p-p_\infty)r 
\,\mathrm{d}r \mathrm{d}\varphi}.
\label{eq:cop}
\ee
Figure~\ref{fig:PDF_2D_CoP} shows probability density functions (PDFs) of the center of pressure for increasing disk inclinations $0^o \leq \theta \leq 10^o$.
Row ($b$) depicts the joint PDF $P(CoP^B_y,CoP^B_z)$ of the CoP location in Cartesian coordinates. 
PDFs for the polar coordinates are shown in rows $(a)$ and $(c)$, respectively,
\tcr{where $A>0$ denotes the amplitude of the wake deflection (radius of the CoP) 
and $\varphi$ its orientation (angle of the CoP)}, 
such that 
\be 
CoP^B_y=A\cos(\varphi),
\quad
CoP^B_z=A\sin(\varphi).
\ee
The PDF of $P(CoP^B_y)$ is also shown in row $(a)$.

When the disk is aligned ($\theta=0$), the base CoP distribution is  axisymmetric. In other words, the disk axisymmetry is recovered statistically, although symmetry is broken instantaneously by the wake's stochastic motion.
This is similar to observations from \cite{ Grandemange_2014-ExiF} and \cite{Rigas_2014-JFM}. A qualitative difference is that the CoP of the disk lies more often in the center, whereas the CoP of the bullet-shaped body lies more often off-center on a circle of finite radius; this is consistent with the absence of antisymmetric wake deflection modes $m=\pm1$ for the disk (section~\ref{sec:base}).

As the disk becomes more inclined ($|\theta|>0$), the base CoP distribution loses symmetry and becomes biased towards one side of the axis, here $y>0$ when $\theta>0$. A similar displacement was observed for small body misalignment by \cite{Gentile-2017}. We note that the asymmetry mainly results from a concentration of the CoP orientation towards $\varphi=0$, while the radius distribution $P(A)$ remains essentially unchanged.

To model this behaviour, we extend the low-order stochastic model of  \cite{Rigas_2014-JFM} as follows:
\begin{equation} 
\dot A = a A -b A^3 + \frac{\Gamma}{A} + \sqrt{2\Gamma} \,\xi_A,
\quad
\dot \varphi = 
-c \theta \sin(\varphi) 
+ \frac{1}{A} \sqrt{2\Gamma} \,\xi_\varphi,
\label{eq:A_Phi_th}
\end{equation}
where 
\tcr{$A$ is the amplitude introduced above, and the overdot denotes time derivative.}
The coefficients $a$ \tcr{(positive or negative)}, and $b$, $c$ and $\Gamma$ 
\tcr{(all positive)} are  to be determined.
The deterministic part of the radius equation is a Landau amplitude equation, $\dot A=a A -b A^3$. 
For a positive linear growth rate $a>0$ (unstable mode), the nonlinear cubic term ensures saturation at a finite equilibrium radius $A=\sqrt{a/b}$. Here we expect a negative growth rate $a<0$ since wake deflection modes are not observed, leading to the deterministic equilibrium $A=0$.
The deterministic part of the angle equation is a restoring force that pushes the orientation towards the stable equilibrium solution $\varphi=0$ ($y>0$) when $\theta>0$, and towards
$\varphi=\pi$ ($y<0$) when $\theta<0$. This restoring term is taken proportional to $\theta$, since it grows stronger (and the CoP distribution gets more asymmetric) as the disk becomes more inclined.
Stochastic forcing induced by the turbulent flow is modeled with independent white Gaussian noises of zero mean and unit intensity, i.e. 
$\left\langle \xi_j(t) \xi_k(t+\tau) \right\rangle 
= \delta_{jk} \delta(\tau)$, with  $j,k=A$ or $\varphi$.
In this model, the radius dynamics are unaffected by the disk inclination, in line with our earlier observation.
As the stochastic forcing constantly deflects the wake, the term $\Gamma/A$ prevents the radius from vanishing and the most probable $A$ is strictly positive (Fig.~\ref{fig:PDF_2D_CoP}$a$). (This does not contradict the most probable $CoP_y^B$ and $CoP_z^B$ being zero, and is a simple consequence of the change of coordinates from Cartesian to polar: $P(A,\varphi) = A \, P(CoP_y^B,CoP_z^B)$.)
When the disk is aligned ($\theta=0$), the angle dynamics are akin to a random walk, but with a radius-dependent effective noise intensity: $\dot\varphi = (1/A) \sqrt{2\Gamma} \,\xi_\varphi$.
When the disk is inclined ($|\theta|>0$), the angle dynamics becomes affected by a non-uniform potential.

The evolution of the PDF $P(A,\varphi)$ is governed by the Fokker-Planck equation
\be
 \dot P = 
-\partial_A                \left( D^{(1)}_{A}              P \right)
-\partial_\varphi          \left( D^{(1)}_{\varphi}        P \right)
+\partial_{AA}             \left( D^{(2)}_{A}             P \right)
+\partial_{\varphi\varphi} \left( D^{(2)}_{\varphi} P \right)
\ee
associated with the stochastic differential equations (\ref{eq:A_Phi_th}), where the drift and diffusion coefficients
(first and second Kramers-Moyal coefficients) \citep{Stratonovich1967, Risken84} read
\be 
D^{(1)}_{A}              = a A -b A^3 + \frac{\Gamma}{A}, \quad
D^{(1)}_{\varphi}        = -c \theta \sin(\varphi), \quad
D^{(2)}_{A}             = \Gamma, \quad
D^{(2)}_{\varphi} = \frac{\Gamma}{A^2}.
\label{eq:FPE}
\ee 
The long-term PDF is obtained under stationary conditions ($\dot P=0$). When the disk is aligned ($\theta=0$), this yields the analytical expressions
\begin{equation}
P(A) 
= \mathcal{N_A} \exp \left[ \frac{1}{\Gamma} 
\left(\frac{a A^2}{2}  -\frac{bA^4}{4} + \Gamma \ln A \right) \right],
\quad
P(\varphi)=\frac{1}{2\pi},
\label{eq:PDF_A_Phi_stat}
\end{equation}
with $\mathcal{N_A}$ a normalization constant such that $\int P(A) \,\mathrm{d}A=1$.
All orientations are statistically equiprobable and $P(\varphi)$ is uniform.
When the disk is inclined ($\theta=0$), a simplified expression is obtained by assuming that the radius remains close to its mean value $A \simeq \Amean$ in the angle dynamics. 
This approximation makes the diffusion coefficient $D^{(2)}_{\varphi}$ independent of $A$ and allows a separation of variables:
\begin{equation}
P(A,\varphi) 
\simeq P(A) P(\varphi)
= \mathcal{N_{A,\varphi}} \exp \left[ \frac{1}{\Gamma} 
\left(\frac{a A^2}{2}  -\frac{bA^4}{4} + \Gamma \ln A \right) \right]
\exp \left[ \frac{\Amean^2}{\Gamma} c \theta \cos(\varphi) \right].
\label{eq:PDF_A_Phi_simple}
\end{equation}
It is interesting to rewrite (\ref{eq:PDF_A_Phi_simple}) as
\begin{equation}
P(A,\varphi) 
\simeq \mathcal{N_{A,\varphi}} 
\exp \left[ -\frac{\mathcal{V_A}}{\Gamma} 
 \right]
\exp \left[ - \frac{\mathcal{V_\varphi}}{\Gamma/\Amean^2} \right],
\end{equation}
using the potentials $\mathcal{V_A}(A)= -a A^2/2 +b A^4/4 -\Gamma \ln A$ and $\mathcal{V_\varphi}(\varphi) = -c \theta \cos(\varphi)$ associated with the stochastic differential equations  such that the deterministic forces in (\ref{eq:A_Phi_th}) derive from the potentials:
$D^{(1)}_{A} = -\mathrm{d}\mathcal{V_A}/\mathrm{d}A$,
$D^{(1)}_{\varphi} = -\mathrm{d}\mathcal{V_\varphi}/\mathrm{d}\varphi$.

\subsection{Parameter identification}

We identify the parameters of the stochastic model (\ref{eq:A_Phi_th}) in two steps: first, we find $a$, $b$ and $\Gamma$ from measurements with the disk fixed and aligned ($\theta=0$); 
second, we find $c$ from measurements with the disk fixed and inclined ($|\theta|>0$).

A fit of the stationary PDF $P(A)$ can only determine the ratios $a/\Gamma$ and $b/\Gamma$, as seen from (\ref{eq:PDF_A_Phi_stat})-(\ref{eq:PDF_A_Phi_simple}). 
In order to determine $a$, $b$ and $\Gamma$ unambiguously, we take advantage of the turbulence-induced stochastic forcing that drives the system to explore the potential landscape $\mathcal{V}(A)$. We use the dynamic information contained in the time signal $A(t)$ and estimate the Kramers-Moyal coefficients \citep{Boettcher06, Friedrich2011}, noting that they can be expressed as 
\be 
D^{(n)}_A(A) = \lim_{\tau\rightarrow 0}   \frac{1}{n!\tau} \int_{0}^{\infty} (A'-A)^n P(A',t+\tau|A,t) \,\mathrm{d}A', 
\quad n=1,2,
\label{eq:D}
\ee
i.e. the short-time limit of transition moments, or  moments of the conditional PDF $P(a,t+\tau|A,t)$ that  describes  the probability of the signal being $A'$ at time $t+\tau$ knowing that it is $A$ at time $t$ \citep{Stratonovich1967, Risken84}.
This method was applied to analyze stochastic data sets in various systems: 
turbulence \citep{Friedrich97PRL, Renner2002}, 
traffic flow \citep{Kriso02}, 
epileptic brains \citep{Prusseit2007},
earthquakes \citep{Manshour2009}
and wind-energy \citep{Milan2013}.
Here we use a version where robustness and accuracy are improved by iteratively minimizing the difference between 
(i)~the moments estimated from time signals with  (\ref{eq:D}),
and 
(ii)~the moments calculated with the adjoint Fokker-Planck equation \citep{LadePLA09, Honisch11, Boujo2017}.
We find 
\be 
a = -11.7 \mbox{ s}^{-1},  \quad
b = 0.22  \mbox{ s}^{-1}, 
\quad  
\Gamma = 6.8\times 10^{-3} \mbox{ s}^{-1}.
\ee
The resulting analytical PDF is shown in Fig.~\ref{fig:PDF_2D_CoP}$(a)$ with its experimental counterpart, illustrating the good agreement obtained.

Next, we use simple fits of $P(\varphi)$ in (\ref{eq:PDF_A_Phi_simple}) at different disk inclinations to find 
\be c=103 \mbox{ s}^{-1}. \ee
Figure \ref{fig:PDF_2D_CoP}$(c)$ shows again a good agreement between the identified and measured PDFs.

\section{Fluid-structure interaction}
\label{sec:FSImodel}

We now move to the fluid-structure interaction problem as  we let the disk free to rotate. Our goal is to derive the simplest  possible model describing the coupled dynamics of the CoP (wake deflection) and the disk (inclination). 
In the previous section, the stochastic equations (\ref{eq:A_Phi_th}) described the dynamics of the CoP, including the influence of the disk inclination.
In this section, we will close the model by describing the dynamics of the disk inclination, and how they are affected by the CoP location.

\subsection{Unsteady aerodynamic loading}

\begin{figure}
\begin{center}
\includegraphics{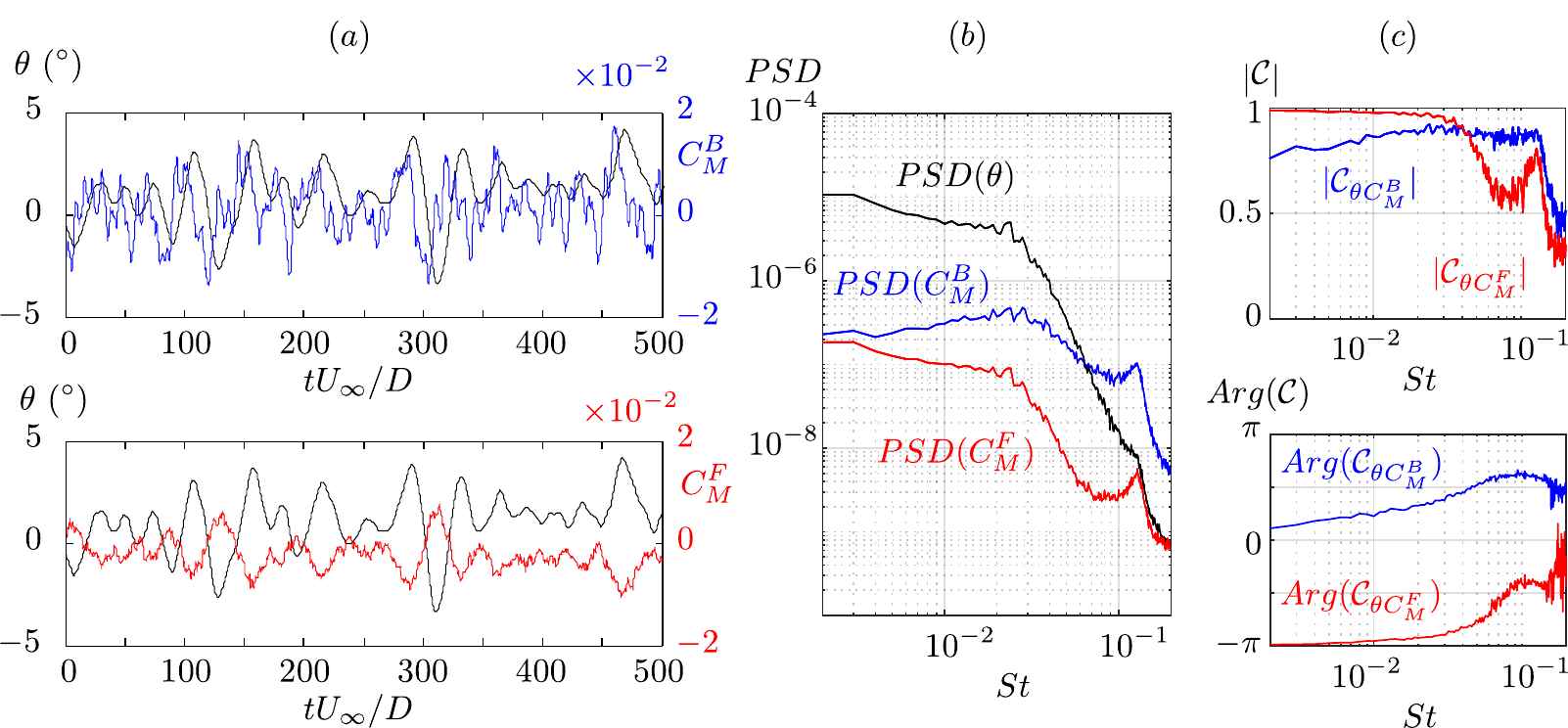}
\end{center}
\caption{
$(a)$~Time signals of the disk inclination and front and base moments.
$(b)$~Power spectral densities.
$(c)$~Coherence functions (magnitude and phase) between inclination and moments.
}
\label{fig:freedisk_exp}
\end{figure}

Experimental results for the moments and disk inclination are shown in Fig.~\ref{fig:freedisk_exp}. 
The time series in Fig.~\ref{fig:freedisk_exp}($a$) are 
low-pass filtered at $St=0.2$.
The shedding frequency at St$\simeq 0.125$ is clearly observable in both moment coefficients spectra in Fig.~\ref{fig:freedisk_exp}($b$) but hardly distinguishable in the angle spectrum. 
The coherence function $\mathcal C$ displayed in magnitude ($|\mathcal C|$) and phase ($Arg(\mathcal C)$) in Fig.~\ref{fig:freedisk_exp}($c$), indicates that the base moment is always in advance compared to the disk inclination while the front moment is delayed and mainly in phase opposition. Notice that the coherence modulus between the angle and the front moment is close to one in the the most energetic frequency domain, emphasizing a strong linear relationship.

To summarize, the dynamics of $\theta$ is forced by the base moment (turbulent wake) while the front pressure loading responds with a restoring moment to the disk inclination. A linear regression from the scatter plot of the front moment vs. the angle gives 
\be  
C_M^F=-0.127 \,\theta.
\label{eq:hydro_stiff2}
\ee
The stiffness is increased by $25\%$ compared to that of the fixed disk in  (\ref{eq:hydro_stiff}), likely due to a structural origin caused by the vinyl tubes of the pressure taps exiting the axis at each of its ends.

\subsection{Model of the disk inclination}

\begin{figure}
\setlength{\unitlength}{1cm}
\begin{center}
   \begin{overpic}[height=7cm, trim=10mm 65mm 20mm 65mm, clip=true]{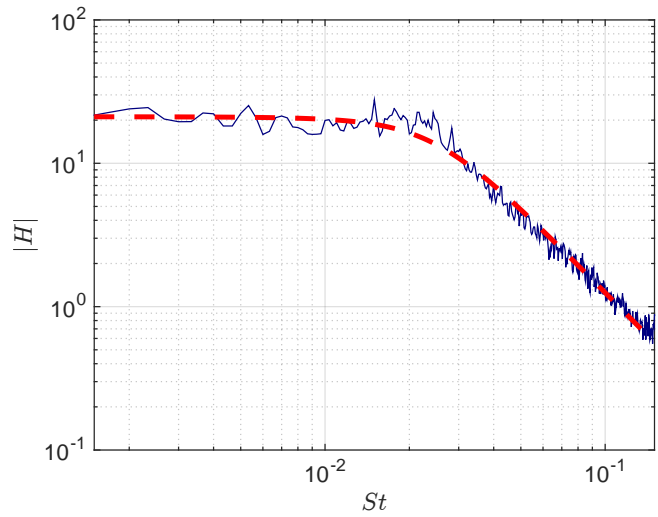}
   \end{overpic}
\end{center}
\vspace{-0.5cm}
\caption{
Gain of the transfer function from  base moment to disk inclination, for the disk free to rotate: (---) measurements and  
(- -) oscillator model~(\ref{eq:theta_oscil_1}).
}
\label{fig:H}
\end{figure}

To gain insight into the inclination dynamics, we compute the gain of the transfer function from  base moment (input) to  inclination (output), using ratios of cross power spectral densities: 
$\left| H(\omega) \right|^2 
=
\left| \mathcal{S}_{\theta\theta} / \mathcal{S}_{\theta M^B} \right|^2
=
\mathcal{S}_{\theta\theta} / \mathcal{S}_{M^B M^B}.$
Figure~\ref{fig:H} shows that the gain has the typical shape of a second-order low-pass system.
Therefore, let us assume that the disk inclination behaves like a linear harmonic oscillator forced by the base moment:
\begin{equation}
\ddot \theta + 2 \alpha \dot \theta + \frac{k}{I} \theta = \frac{M^B}{I}.
\label{eq:theta_oscil_1}
\end{equation}
In this equation, $I$ is the disk's moment of inertia about the rotation axis $(Oz)$, $k$ is the aerodynamic stiffness resulting from the font face loading as defined in (\ref{eq:hydro_stiff})-(\ref{eq:hydro_stiff2}), and $\alpha$ is a damping coefficient to be determined.
Defining the cut-off frequency $\omega_0=\sqrt{k/I}$ and fitting the analytical expression of the gain 
\be 
|H(\omega)|^2
= \left|\frac{1/I}{ (i\omega)^2 + 2 \alpha i\omega  + \omega_0^2 }\right|^2 
= \frac{ 1/I^2 }{ (\omega_0^2-\omega^2)^2 + (2 \alpha \omega)^2 }
\label{eq:gain}
\ee 
yields the parameters
\be 
I = 8.0 \times 10^{-4} \mbox{ kg.m}^2, 
\quad
\alpha = 5.9 \mbox{ s}^{-1}, 
\quad  
\omega_0/(2\pi) = 1.22 \mbox{ Hz } (St_0=0.024), 
\label{eq:fit}
\ee
and a good agreement with experimental data (Fig.~\ref{fig:H}).
We note that the identified value of $I$ 
corresponds to an equivalent density  $\rho=400$~kg/m$^3$  for a homogeneous disk ($I = \rho \iiint (x^2+y^2) \mathrm{d}x \mathrm{d}y \mathrm{d}z= 2\times 10^{-6} \rho$~kg.m$^2$ with the present disk's dimensions), which is consistent with the disk's material and geometry.
We also note that the identified values of $\omega_0$ and $I$ yield  $k/(qSD) = \omega_0^2 I /(qSD) = 0.121$, not far from the aerodynamic stiffness values reported in (\ref{eq:hydro_stiff})-(\ref{eq:hydro_stiff2}), determined from static and dynamic measurements of the front moment, respectively.

Next, we express the base moment as a function of the CoP location, 
\be M^B=F_N^B 
 R A  \cos(\varphi),
\label{eq:moment_from_A_Phi}
\ee
and we introduce the angular velocity $\Omega=\dot\theta$ to recast the oscillator with first-order equations.
Finally, we obtain a closed model for the variables $A, \varphi, \theta$ and $\dot\theta$:
\begin{align} 
\dot A &= a A -b A^3 + \frac{\Gamma}{A} + \sqrt{2\Gamma} \,\xi_A,
\label{eq:A_Phi_th_thdot_1}
\\
\dot\varphi &= -c \theta \sin(\varphi) + \frac{1}{A} \sqrt{2\Gamma} \,\xi_\varphi,
\label{eq:A_Phi_th_thdot_2}
\\
\dot\theta &= \Omega,
\label{eq:A_Phi_th_thdot_3}
\\
\dot\Omega &= 
- 2\alpha \Omega - \omega_0^2 \theta + \dfrac{F_N^B R}{ I}  \, A \cos(\varphi).
\label{eq:A_Phi_th_thdot_4}
\end{align}

\section{Results and discussions}
\label{sec:results}

We perform time simulations of the stochastic model (\ref{eq:A_Phi_th_thdot_1})-(\ref{eq:A_Phi_th_thdot_4}) for $T=1000$~s ($T U_\infty/D=5\times10^4$) with a time step $\mathrm{d}t=10^{-5}$~s and with the parameters identified in sections \ref{sec:fixedmodel}-\ref{sec:FSImodel}.

\begin{figure}
\setlength{\unitlength}{1cm}
\begin{center}
  \begin{overpic}[height=8cm, trim=12mm 65mm 20mm 70mm, clip=true]{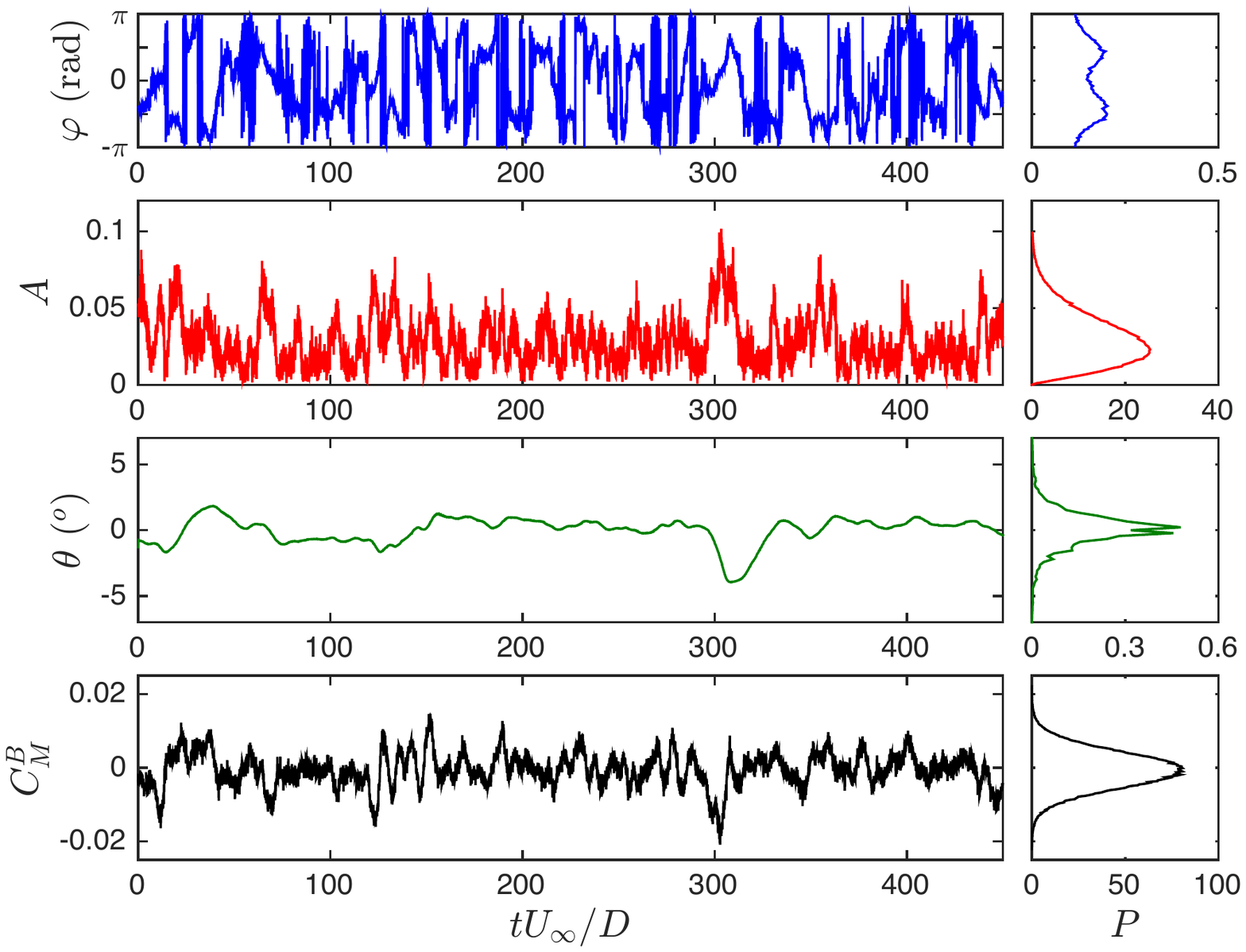}
      \put(-8,74){$(a)$}
   \end{overpic}
   \begin{overpic}[height=8cm, trim=12mm 65mm 20mm 70mm, clip=true]{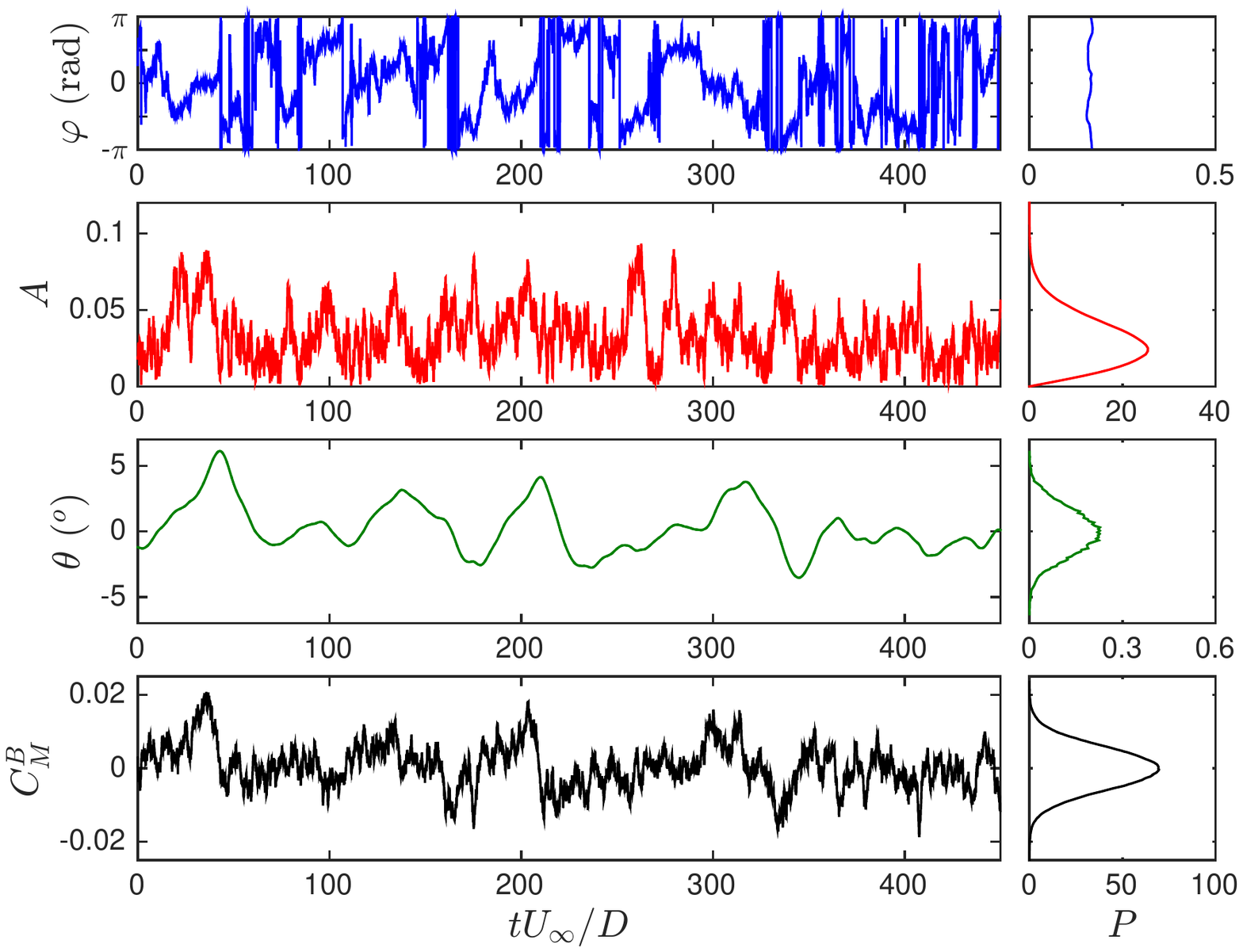}
      \put(-8,74){$(b)$}
   \end{overpic}
\end{center}
\vspace{-0.7cm}
\caption{
Time series and PDF of the CoP orientation, CoP radius, 
disk inclination and base moment coefficient.
Disk free to rotate.
$(a)$ Measurements,
$(b)$ stochastic model.
}
\label{fig:PDFs}
\end{figure}

Figure~\ref{fig:PDFs} shows simulation results together with experimental measurements. We observe a good overall agreement, both dynamic (time series) and statistic (PDFs), for all the relevant variables: from base CoP location to disk inclination to base moment.
In particular, the statistical axisymmetry of the CoP is recovered, as well as its radius distribution, which validates the stochastic model (\ref{eq:A_Phi_th}) for the CoP dynamics, including the influence of the disk inclination.
The base moment's characteristics are well captured too, which validates the simple relation (\ref{eq:moment_from_A_Phi}). Note that we found no significant difference in the results when using either
(i)~the mean value of the force 
$F_N^B = \left\langle F_N^B \right\rangle = 0.91$~N,
or 
(ii)~the experimental Gaussian distribution $F_N^B(t) = \left\langle F_N^B \right\rangle + F_N^{B'}(t)$ 
with standard deviation 
$\langle (F_N^{B'})^2 \rangle^{1/2} = 0.09$~N.
Regarding the disk inclination, although the slow dynamics are well captured,
some discrepancy is observed in the PDF: the experimental $P(\theta)$ has exponential tails whereas the stochastic model yields a Gaussian $P(\theta)$.
This means that the actual disk exhibits ``extreme events'' (rare but large deviations from $\langle \theta \rangle=0$) that are not accounted for in the model. This suggests modifying the simple linear harmonic oscillator (\ref{eq:theta_oscil_1}), for example by adding non-linearities in $\theta$, or by  including time delay effects between $\theta$ and the base moment $M^B$  (recall Fig.~\ref{fig:freedisk_exp}).

We note that the stochastic model (\ref{eq:A_Phi_th_thdot_1})-(\ref{eq:A_Phi_th_thdot_4}) could easily be recast with the base CoP location in Cartesian coordinates, using the change of variables 
$CoP_y^B=A\cos(\varphi)$, $CoP_z^B=A\sin(\varphi)$.
Alternatively, it could be recast with the base moments about the two axes $(Oz)$ and $(Oy)$, because moment-to-force ratios are proportional to the CoP Cartesian coordinates (e.g. $M^B/F^B\propto CoP_y^B$ as seen from (\ref{eq:def_moment_force_cop}) and (\ref{eq:cop}), with a similar relation for $CoP_z^B$).

\section{Conclusion}
\label{sec:conclusion}

We investigate the fundamental problem of coupled interaction between the inclination of a bluff body and its turbulent wake.
We consider a thin disk free to rotate and perform time-resolved measurements of front and base pressure and disk inclination $\theta$ in a wind tunnel at $Re\simeq 1.33 \times 10^5$.

When the disk is fixed and aligned, we  find low-frequency axisymmetric ($m=0$) wake pumping and asymmetric ($m=\pm1$) vortex shedding at $St \simeq 0.125$. 
Very-low-frequency antisymmetric $m=\pm1$  modes are not observed, unless the disk is equipped with a hemispheric forebody and thus transformed into a bullet shape similar to that of \cite{Grandemange_2014-ExiF, Rigas_2014-JFM, Rigas_2015-JFM, Rigas_2016, Gentile-2016, Gentile-2017}.
The instantaneous symmetry breaking vortex shedding visits all orientations with equal probability, yielding an axisymmetric long-term distribution of the base center of pressure, similar to 
 \cite{ Grandemange_2014-ExiF} and \cite{Rigas_2014-JFM}. 
 
When the disk is fixed and inclined, 
 the base CoP distribution moves preferentially towards one side of the rotation axis, similar to observations by \cite{Gentile-2017}. The front moment is proportional to $\theta$ and corresponds to an aerodynamic stiffness.

When the disk is free to rotate, the front moment is in phase opposition with $\theta$, confirming its static role as an aerodynamic stiffness; the base moment shows a good coherence with $\theta$, although the disk inclination lags behind.

Based on those observations, we build the following coupled stochastic model:
(i)~the CoP location is governed by stochastic dynamics forced by the turbulent flow; the orientation visits in a random walk-like fashion a potential $\mathcal{V}(\varphi)$ that is minimum in $\varphi=0$ for positive inclinations $\theta>0$ and minimum in $\varphi=\pi$ for negative inclinations $\theta<0$; 
the radius visits a potential $\mathcal{V}(A)$ that is infinite in $A=0$ and minimum in a preferred $A>0$, even though the mode is stable, due to the stochastic forcing;
(ii)~the disk inclination is a linear harmonic oscillator forced by the base moment.
The model captures well the main statistic and dynamics features observed experimentally. One notable difference lies in the distribution $P(\theta)$ of the disk inclination, which features exponential tails in the measurements, suggesting a possible refinement of the model with non-linearities or time delays.

We expect our low-order stochastic model to capture coupled dynamics in other fluid-structure interaction systems involving turbulent wakes behind bluff bodies free to rotate, provided that the free parameters are determined correctly.
In particular, very-low-frequency symmetry-breaking modes observed in the wake of bullet shapes but not in the wake of the disk can be ``turned on'' or ``turned off'' simply by changing their linear stability (i.e. the sign of a single parameter).
Non-axisymmetric bluff bodies such as rectangular plates or prisms could be handled using a similar model with minor modifications, e.g.  polar coordinates describing an ellipse rather than a circle \citep{Bonnavion2018jfm}.

\section*{Acknowledgements}

E.B. thanks Thomas Leweke and the organising committee of the 7$^{th}$ Conference on Bluff Body Wakes and Vortex-Induced Vibrations (BBVIV-7)  for making his participation possible. 
O.C. is grateful to Sebastiano Fischera for useful discussions and John Curran from the workshop of the University of Liverpool for building and designing the models.

\section*{References}

\bibliography{mybibfile}

\begin{thebibliography}{29}
\expandafter\ifx\csname natexlab\endcsname\relax\def\natexlab#1{#1}\fi
\providecommand{\url}[1]{\texttt{#1}}
\providecommand{\href}[2]{#2}
\providecommand{\path}[1]{#1}
\providecommand{\DOIprefix}{doi:}
\providecommand{\ArXivprefix}{arXiv:}
\providecommand{\URLprefix}{URL: }
\providecommand{\Pubmedprefix}{pmid:}
\providecommand{\doi}[1]{\href{http://dx.doi.org/#1}{\path{#1}}}
\providecommand{\Pubmed}[1]{\href{pmid:#1}{\path{#1}}}
\providecommand{\bibinfo}[2]{#2}
\ifx\xfnm\relax \def\xfnm[#1]{\unskip,\space#1}\fi
\bibitem[{Achenbach(1974)}]{Achenbach-1974}
\bibinfo{author}{Achenbach, E.}, \bibinfo{year}{1974}.
\newblock \bibinfo{title}{{Vortex shedding from spheres}}.
\newblock \bibinfo{journal}{Journal of Fluid Mechanics} \bibinfo{volume}{62},
  \bibinfo{pages}{209--221}.
\bibitem[{Berger et~al.(1990)Berger, Scholz and Schumm}]{Berger-1990}
\bibinfo{author}{Berger, E.}, \bibinfo{author}{Scholz, D.},
  \bibinfo{author}{Schumm, M.}, \bibinfo{year}{1990}.
\newblock \bibinfo{title}{{Coherent vortex structures in the wake of a sphere
  and a circular disk at rest and under forced vibrations}}.
\newblock \bibinfo{journal}{Journal of Fluids and Structures}
  \bibinfo{volume}{4}, \bibinfo{pages}{231--257}.
\bibitem[{Bonnavion and Cadot(2018)}]{Bonnavion2018jfm}
\bibinfo{author}{Bonnavion, G.}, \bibinfo{author}{Cadot, O.},
  \bibinfo{year}{2018}.
\newblock \bibinfo{title}{Unstable wake dynamics of rectangular flat-backed
  bluff bodies with inclination and ground proximity}.
\newblock \bibinfo{journal}{Journal of Fluid Mechanics} \bibinfo{volume}{854},
  \bibinfo{pages}{196–232}.
\newblock \DOIprefix\doi{10.1017/jfm.2018.630}.
\bibitem[{B\"ottcher et~al.(2006)B\"ottcher, Peinke, Kleinhans, Friedrich, Lind
  and Haase}]{Boettcher06}
\bibinfo{author}{B\"ottcher, F.}, \bibinfo{author}{Peinke, J.},
  \bibinfo{author}{Kleinhans, D.}, \bibinfo{author}{Friedrich, R.},
  \bibinfo{author}{Lind, P.G.}, \bibinfo{author}{Haase, M.},
  \bibinfo{year}{2006}.
\newblock \bibinfo{title}{Reconstruction of complex dynamical systems affected
  by strong measurement noise}.
\newblock \bibinfo{journal}{Phys. Rev. Lett.} \bibinfo{volume}{97},
  \bibinfo{pages}{090603}.
\newblock \URLprefix
  \url{http://link.aps.org/doi/10.1103/PhysRevLett.97.090603},
  \DOIprefix\doi{10.1103/PhysRevLett.97.090603}.
\bibitem[{Boujo and Noiray(2017)}]{Boujo2017}
\bibinfo{author}{Boujo, E.}, \bibinfo{author}{Noiray, N.},
  \bibinfo{year}{2017}.
\newblock \bibinfo{title}{Robust identification of harmonic oscillator
  parameters using the adjoint {F}okker-{P}lanck equation}.
\newblock \bibinfo{journal}{Proceedings of the Royal Society A: Mathematical,
  Physical and Engineering Science} \bibinfo{volume}{473}.
\newblock \URLprefix
  \url{http://rspa.royalsocietypublishing.org/content/473/2200/20160894.abstract}.
\bibitem[{Cadot(2016)}]{Cadot_2016_jfmrapids}
\bibinfo{author}{Cadot, O.}, \bibinfo{year}{2016}.
\newblock \bibinfo{title}{Stochastic fluid structure interaction of
  three-dimensional plates facing a uniform flow}.
\newblock \bibinfo{journal}{Journal of Fluid Mechanics} \bibinfo{volume}{794}.
\bibitem[{Fabre et~al.(2008)Fabre, Auguste and Magnaudet}]{Fabre_2008}
\bibinfo{author}{Fabre, D.}, \bibinfo{author}{Auguste, F.},
  \bibinfo{author}{Magnaudet, J.}, \bibinfo{year}{2008}.
\newblock \bibinfo{title}{Bifurcations and symmetry breaking in the wake of
  axisymmetric bodies}.
\newblock \bibinfo{journal}{Physics of Fluids} \bibinfo{volume}{20},
  \bibinfo{pages}{051702}.
\bibitem[{Friedrich and Peinke(1997)}]{Friedrich97PRL}
\bibinfo{author}{Friedrich, R.}, \bibinfo{author}{Peinke, J.},
  \bibinfo{year}{1997}.
\newblock \bibinfo{title}{Description of a turbulent cascade by a
  {F}okker-{P}lanck equation}.
\newblock \bibinfo{journal}{Phys. Rev. Lett.} \bibinfo{volume}{78},
  \bibinfo{pages}{863--866}.
\newblock \URLprefix \url{http://link.aps.org/doi/10.1103/PhysRevLett.78.863},
  \DOIprefix\doi{10.1103/PhysRevLett.78.863}.
\bibitem[{Friedrich et~al.(2011)Friedrich, Peinke, Sahimi and
  Tabar}]{Friedrich2011}
\bibinfo{author}{Friedrich, R.}, \bibinfo{author}{Peinke, J.},
  \bibinfo{author}{Sahimi, M.}, \bibinfo{author}{Tabar, M.R.R.},
  \bibinfo{year}{2011}.
\newblock \bibinfo{title}{Approaching complexity by stochastic methods: From
  biological systems to turbulence}.
\newblock \bibinfo{journal}{Physics Reports} \bibinfo{volume}{506},
  \bibinfo{pages}{87--162}.
\bibitem[{Fuchs et~al.(1979)Fuchs, Mercker and
  Michel}]{fuchs_mercker_michel_1979}
\bibinfo{author}{Fuchs, H.V.}, \bibinfo{author}{Mercker, E.},
  \bibinfo{author}{Michel, U.}, \bibinfo{year}{1979}.
\newblock \bibinfo{title}{Large-scale coherent structures in the wake of
  axisymmetric bodies}.
\newblock \bibinfo{journal}{Journal of Fluid Mechanics} \bibinfo{volume}{93},
  \bibinfo{pages}{185–207}.
\bibitem[{Gentile et~al.(2016)Gentile, Schrijer, Van~Oudheusden and
  Scarano}]{Gentile-2016}
\bibinfo{author}{Gentile, V.}, \bibinfo{author}{Schrijer, F.},
  \bibinfo{author}{Van~Oudheusden, B.}, \bibinfo{author}{Scarano, F.},
  \bibinfo{year}{2016}.
\newblock \bibinfo{title}{Low-frequency behavior of the turbulent axisymmetric
  near-wake}.
\newblock \bibinfo{journal}{Physics of Fluids} \bibinfo{volume}{28}.
\bibitem[{Gentile et~al.(2017)Gentile, Van~Oudheusden, Schrijer and
  Scarano}]{Gentile-2017}
\bibinfo{author}{Gentile, V.}, \bibinfo{author}{Van~Oudheusden, B.},
  \bibinfo{author}{Schrijer, F.}, \bibinfo{author}{Scarano, F.},
  \bibinfo{year}{2017}.
\newblock \bibinfo{title}{The effect of angular misalignment on low-frequency
  axisymmetric wake instability}.
\newblock \bibinfo{journal}{Journal of Fluid Mechanics} \bibinfo{volume}{813}.
\bibitem[{Grandemange et~al.(2014)Grandemange, Gohlke and
  Cadot}]{Grandemange_2014-ExiF}
\bibinfo{author}{Grandemange, M.}, \bibinfo{author}{Gohlke, M.},
  \bibinfo{author}{Cadot, O.}, \bibinfo{year}{2014}.
\newblock \bibinfo{title}{{Statistical axisymmetry of the turbulent sphere
  wake}}.
\newblock \bibinfo{journal}{Experiments in fluids} \bibinfo{volume}{55},
  \bibinfo{pages}{1--10}.
\bibitem[{Honisch and Friedrich(2011)}]{Honisch11}
\bibinfo{author}{Honisch, C.}, \bibinfo{author}{Friedrich, R.},
  \bibinfo{year}{2011}.
\newblock \bibinfo{title}{Estimation of {K}ramers-{M}oyal coefficients at low
  sampling rates}.
\newblock \bibinfo{journal}{Physical Review E} \bibinfo{volume}{83},
  \bibinfo{pages}{066701}.
\newblock \URLprefix \url{http://link.aps.org/doi/10.1103/PhysRevE.83.066701},
  \DOIprefix\doi{10.1103/PhysRevE.83.066701}.
\bibitem[{Kriso et~al.(2002)Kriso, Peinke, Friedrich and Wagner}]{Kriso02}
\bibinfo{author}{Kriso, S.}, \bibinfo{author}{Peinke, J.},
  \bibinfo{author}{Friedrich, R.}, \bibinfo{author}{Wagner, P.},
  \bibinfo{year}{2002}.
\newblock \bibinfo{title}{Reconstruction of dynamical equations for traffic
  flow}.
\newblock \bibinfo{journal}{Physics Letters A} \bibinfo{volume}{299},
  \bibinfo{pages}{287 -- 291}.
\newblock \URLprefix
  \url{http://www.sciencedirect.com/science/article/pii/S0375960102002888},
  \DOIprefix\doi{http://dx.doi.org/10.1016/S0375-9601(02)00288-8}.
\bibitem[{Lade(2009)}]{LadePLA09}
\bibinfo{author}{Lade, S.}, \bibinfo{year}{2009}.
\newblock \bibinfo{title}{Finite sampling interval effects in {K}ramers-{M}oyal
  analysis}.
\newblock \bibinfo{journal}{Physics Letters A} \bibinfo{volume}{373},
  \bibinfo{pages}{3705--3709}.
\bibitem[{Manshour et~al.(2009)Manshour, Saberi, Sahimi, Peinke, Pacheco and
  Rahimi~Tabar}]{Manshour2009}
\bibinfo{author}{Manshour, P.}, \bibinfo{author}{Saberi, S.},
  \bibinfo{author}{Sahimi, M.}, \bibinfo{author}{Peinke, J.},
  \bibinfo{author}{Pacheco, A.F.}, \bibinfo{author}{Rahimi~Tabar, M.R.},
  \bibinfo{year}{2009}.
\newblock \bibinfo{title}{Turbulencelike behavior of seismic time series}.
\newblock \bibinfo{journal}{Phys. Rev. Lett.} \bibinfo{volume}{102},
  \bibinfo{pages}{014101}.
\newblock \URLprefix
  \url{https://link.aps.org/doi/10.1103/PhysRevLett.102.014101}.
\bibitem[{Meliga et~al.(2009)Meliga, Chomaz and Sipp}]{Meliga_2009}
\bibinfo{author}{Meliga, P.}, \bibinfo{author}{Chomaz, J.},
  \bibinfo{author}{Sipp, D.}, \bibinfo{year}{2009}.
\newblock \bibinfo{title}{{Unsteadiness in the wake of disks and spheres:
  instability, receptivity and control using direct and adjoint global
  stability analyses}}.
\newblock \bibinfo{journal}{Journal of Fluids and Structures}
  \bibinfo{volume}{25}, \bibinfo{pages}{601--616}.
\bibitem[{Milan et~al.(2013)Milan, W\"achter and Peinke}]{Milan2013}
\bibinfo{author}{Milan, P.}, \bibinfo{author}{W\"achter, M.},
  \bibinfo{author}{Peinke, J.}, \bibinfo{year}{2013}.
\newblock \bibinfo{title}{Turbulent character of wind energy}.
\newblock \bibinfo{journal}{Phys. Rev. Lett.} \bibinfo{volume}{110},
  \bibinfo{pages}{138701}.
\newblock \URLprefix
  \url{https://link.aps.org/doi/10.1103/PhysRevLett.110.138701}.
\bibitem[{Prusseit and Lehnertz(2007)}]{Prusseit2007}
\bibinfo{author}{Prusseit, J.}, \bibinfo{author}{Lehnertz, K.},
  \bibinfo{year}{2007}.
\newblock \bibinfo{title}{Stochastic qualifiers of epileptic brain dynamics}.
\newblock \bibinfo{journal}{Phys. Rev. Lett.} \bibinfo{volume}{98},
  \bibinfo{pages}{138103}.
\newblock \URLprefix
  \url{https://link.aps.org/doi/10.1103/PhysRevLett.98.138103},
  \DOIprefix\doi{10.1103/PhysRevLett.98.138103}.
\bibitem[{Renner et~al.(2002)Renner, Peinke, Friedrich, Chanal and
  Chabaud}]{Renner2002}
\bibinfo{author}{Renner, C.}, \bibinfo{author}{Peinke, J.},
  \bibinfo{author}{Friedrich, R.}, \bibinfo{author}{Chanal, O.},
  \bibinfo{author}{Chabaud, B.}, \bibinfo{year}{2002}.
\newblock \bibinfo{title}{Universality of small scale turbulence}.
\newblock \bibinfo{journal}{Phys. Rev. Lett.} \bibinfo{volume}{89},
  \bibinfo{pages}{124502}.
\newblock \URLprefix
  \url{https://link.aps.org/doi/10.1103/PhysRevLett.89.124502}.
\bibitem[{Rigas et~al.(2016)Rigas, Esclapez and Magri}]{Rigas_2016}
\bibinfo{author}{Rigas, G.}, \bibinfo{author}{Esclapez, L.},
  \bibinfo{author}{Magri, L.}, \bibinfo{year}{2016}.
\newblock \bibinfo{title}{{Symmetry breaking in a 3D bluff-body wake}},
  \bibinfo{organization}{Center for Turbulence Research, Stanford University}.
\bibitem[{Rigas et~al.(2015)Rigas, Morgans, Brackston and
  Morrison}]{Rigas_2015-JFM}
\bibinfo{author}{Rigas, G.}, \bibinfo{author}{Morgans, A.},
  \bibinfo{author}{Brackston, R.D.}, \bibinfo{author}{Morrison, J.},
  \bibinfo{year}{2015}.
\newblock \bibinfo{title}{Diffusive dynamics and stochastic models of turbulent
  axisymmetric wakes}.
\newblock \bibinfo{journal}{Journal of Fluid Mechanics} \bibinfo{volume}{778},
  \bibinfo{pages}{R2}.
\bibitem[{Rigas et~al.(2014)Rigas, Oxlade, Morgans and
  Morrison}]{Rigas_2014-JFM}
\bibinfo{author}{Rigas, G.}, \bibinfo{author}{Oxlade, A.},
  \bibinfo{author}{Morgans, A.}, \bibinfo{author}{Morrison, J.},
  \bibinfo{year}{2014}.
\newblock \bibinfo{title}{Low-dimensional dynamics of a turbulent axisymmetric
  wake}.
\newblock \bibinfo{journal}{Journal of Fluid Mechanics} \bibinfo{volume}{755},
  \bibinfo{pages}{159}.
\bibitem[{Risken(1984)}]{Risken84}
\bibinfo{author}{Risken, H.}, \bibinfo{year}{1984}.
\newblock \bibinfo{title}{The {F}okker--{P}lanck Equation}.
\newblock \bibinfo{publisher}{Springer-Verlag}.
\bibitem[{Sakamoto and Haniu(1990)}]{Sakamoto-1990}
\bibinfo{author}{Sakamoto, H.}, \bibinfo{author}{Haniu, H.},
  \bibinfo{year}{1990}.
\newblock \bibinfo{title}{{A study on vortex shedding from spheres in a uniform
  flow}}.
\newblock \bibinfo{journal}{Journal of Fluids Engineering}
  \bibinfo{volume}{112}, \bibinfo{pages}{386--392}.
\bibitem[{Stratonovich(1967)}]{Stratonovich1967}
\bibinfo{author}{Stratonovich, R.}, \bibinfo{year}{1967}.
\newblock \bibinfo{title}{Topics in the Theory of Random Noise}.
  volume~\bibinfo{volume}{2}.
\newblock \bibinfo{publisher}{New York: Gordon \& Breach}.
\bibitem[{Taneda(1978)}]{Taneda-1978}
\bibinfo{author}{Taneda, S.}, \bibinfo{year}{1978}.
\newblock \bibinfo{title}{{Visual observations of the flow past a sphere at
  Reynolds numbers between $10^4$ and $10^6$}}.
\newblock \bibinfo{journal}{Journal of Fluid Mechanics} \bibinfo{volume}{85},
  \bibinfo{pages}{187--192}.
\bibitem[{Yun et~al.(2006)Yun, Kim and Choi}]{Yun-2006}
\bibinfo{author}{Yun, G.}, \bibinfo{author}{Kim, D.}, \bibinfo{author}{Choi,
  H.}, \bibinfo{year}{2006}.
\newblock \bibinfo{title}{{Vortical structures behind a sphere at subcritical
  Reynolds numbers}}.
\newblock \bibinfo{journal}{Physics of Fluids} \bibinfo{volume}{18},
  \bibinfo{pages}{015102}.

\end{thebibliography}

\end{document}